\documentclass[]{aastex701}

\begin{document}

\title{Predicting Radial Velocities from Rossiter-McLaughlin Time Series Observations}

\author[orcid=0000-0002-0601-6199]{Vardan Adibekyan}
\affiliation{Instituto de Astrof\'isica e Ci\^encias do Espa\c{c}o, Universidade do Porto, CAUP, Rua das Estrelas, 4150-762 Porto, Portugal}
\affiliation{Departamento de F\'{\i}sica e Astronomia, Faculdade de Ci\^encias, Universidade do Porto, Rua do Campo  Alegre, 4169-007 Porto, Portugal}
\email[show]{vadibekyan@astro.up.pt}

\author[orcid=0000-0001-7918-0355]{Olivier Demangeon}
\affiliation{Instituto de Astrof\'isica e Ci\^encias do Espa\c{c}o, Universidade do Porto, CAUP, Rua das Estrelas, 4150-762 Porto, Portugal}
\affiliation{Departamento de F\'{\i}sica e Astronomia, Faculdade de Ci\^encias, Universidade do Porto, Rua do Campo Alegre, 4169-007 Porto, Portugal}
\email{olivier.demangeon@astro.up.pt}

\author[orcid=0009-0009-1511-053X]{Diogo Teixeira}
\affiliation{Faculdade de Ci\^encias, Universidade do Porto, Rua do Campo Alegre, 4169-007 Porto, Portugal}
\email{diogoteixeira1999@gmail.com}

\author[orcid=0000-0003-4422-2919]{Nuno Santos}
\affiliation{Instituto de Astrof\'isica e Ci\^encias do Espa\c{c}o, Universidade do Porto, CAUP, Rua das Estrelas, 4150-762 Porto, Portugal}
\affiliation{Departamento de F\'{\i}sica e Astronomia, Faculdade de Ci\^encias, Universidade do Porto, Rua do Campo Alegre, 4169-007 Porto, Portugal}
\email{nuno@astro.up.pt}

\author[orcid=0000-0002-3212-5778]{Khaled Al Moulla}
\affiliation{Instituto de Astrof\'isica e Ci\^encias do Espa\c{c}o, Universidade do Porto, CAUP, Rua das Estrelas, 4150-762 Porto, Portugal}
\email{khaled.almoulla@astro.up.pt}

\author[orcid=0000-0001-5992-7589]{Eduardo Cristo}
\affiliation{Instituto de Astrof\'isica e Ci\^encias do Espa\c{c}o, Universidade do Porto, CAUP, Rua das Estrelas, 4150-762 Porto, Portugal}
\email{eduardo.cristo@astro.up.pt}

\author[orcid=0000-0003-4920-738X]{André Silva}
\affiliation{Instituto de Astrof\'isica e Ci\^encias do Espa\c{c}o, Universidade do Porto, CAUP, Rua das Estrelas, 4150-762 Porto, Portugal}
\affiliation{Departamento de F\'{\i}sica e Astronomia, Faculdade de Ci\^encias, Universidade do Porto, Rua do Campo Alegre, 4169-007 Porto, Portugal}
\email{Andre.Silva@astro.up.pt}

\author[orcid=0000-0002-5179-4344]{Roman Chertovskih}
\affiliation{Research Center for Systems and Technologies (SYSTEC/ARISE), Faculdade de Engenharia, Universidade do Porto, Rua Dr. Roberto Frias, 4200-465 Porto, Portugal}
\email{roman@fe.up.pt}

\author[]{Garik Israelian}
\affiliation{Instituto de Astrof\'{i}sica de Canarias, E-38205 La Laguna, Tenerife, Spain}
\affiliation{Departamento de Astrof\`{i}sica, Universidad de La Laguna, E-38206 La Laguna, Tenerife, Spain}
\email{gil@iac.es}

\author[orcid=0000-0001-7392-1765]{Artur Hakobyan}
\affiliation{Center for Cosmology and Astrophysics, Alikhanian National Science Laboratory, 2 Alikhanian Brothers Str., 0036 Yerevan, Armenia}
\email{artur.hakobyan@yerphi.am}


\begin{abstract}

The Rossiter--McLaughlin (RM) effect produces apparent radial velocity (RV) shifts through line-profile distortions caused by a transiting planet blocking different regions of the rotating stellar surface. Because the underlying orbital RV trend can be estimated from out-of-transit observations, RM sequences provide a controlled laboratory for studying flux-induced RV variations. We compiled a sample of 1171 ESPRESSO observations of 13 targets obtained during 21 RM observing nights and trained machine-learning models to reconstruct a reference RV trend from observed RVs, line-profile diagnostics, and activity indicators. Predictive performance varied substantially among stars and depended on both the strength of the RM signal and the similarity of the target star to the training sample. Applications to Sun-as-a-star observations and Proxima Centauri recovered known periodicities but did not fully remove the activity-induced variability. Larger and more diverse datasets will be required to assess the potential of this approach for mitigating activity-induced RV signals.
\end{abstract}



\section{Introduction}

Radial velocity (RV) variations induced by orbiting planets arise from Doppler shifts that translate the stellar spectrum without altering its intrinsic line shapes. In contrast, stellar surface inhomogeneities may modify the flux contribution from different regions of the rotating stellar disk, distorting the integrated line profiles and generating apparent RV shifts \citep[e.g.][]{Santos-14}. A similar phenomenon occurs during planetary transits through the Rossiter--McLaughlin (RM) effect, where the planet blocks different regions of the stellar surface and produces a characteristic RV anomaly through line-profile deformation \citep[e.g.][]{Gaudi-07}.

Recent years have seen growing interest in the application of machine-learning (ML) techniques to RV analysis \citep[e.g.][]{McWilliam-26}. Unlike the case of stellar activity, where the uncontaminated RV signal is unknown, RM observations contain out-of-transit measurements that can be used to estimate the orbital RV trend that would be observed in the absence of the RM effect (Figure~\ref{fig:rm_rv_fig}). RM observations therefore allow to study how line-profile distortions translate into apparent RV shifts.  We investigate whether ML models can recover the underlying RV trend from observed RVs and line-profile diagnostics.

\section{Sample and RV data}

We assembled a sample of RM time-series observations from the DACE platform\footnote{\url{https://dace.unige.ch/}}, using  cross-correlation functions (CCFs) from ESPRESSO observations of systems with known RM signals. The CCFs were analysed using \texttt{iCCF}\footnote{\url{https://github.com/j-faria/iCCF}}. For each exposure, we extracted the RV together with line-profile diagnostics (full width at half maximum (FWHM), bisector inverse slope (BIS), Vspan, Wspan, and contrast \citep[see e.g.][]{Santerne-15}). We also extracted  activity indicators based on the Ca, Na, and H$\alpha$ lines \citep[e.g.][]{JGdS-18}.

The initial sample contained 55 observing nights for 41 stars. We visually inspected all RM sequences and retained only observations with a well-sampled RM signal, including out-of-transit measurements before and after transit to define the underlying RV trend. The final sample used for the analysis consists of 1171 exposures obtained during 21 observing nights for 13 stars.

\begin{figure}[htbp]
\centering
\includegraphics[width=\linewidth]{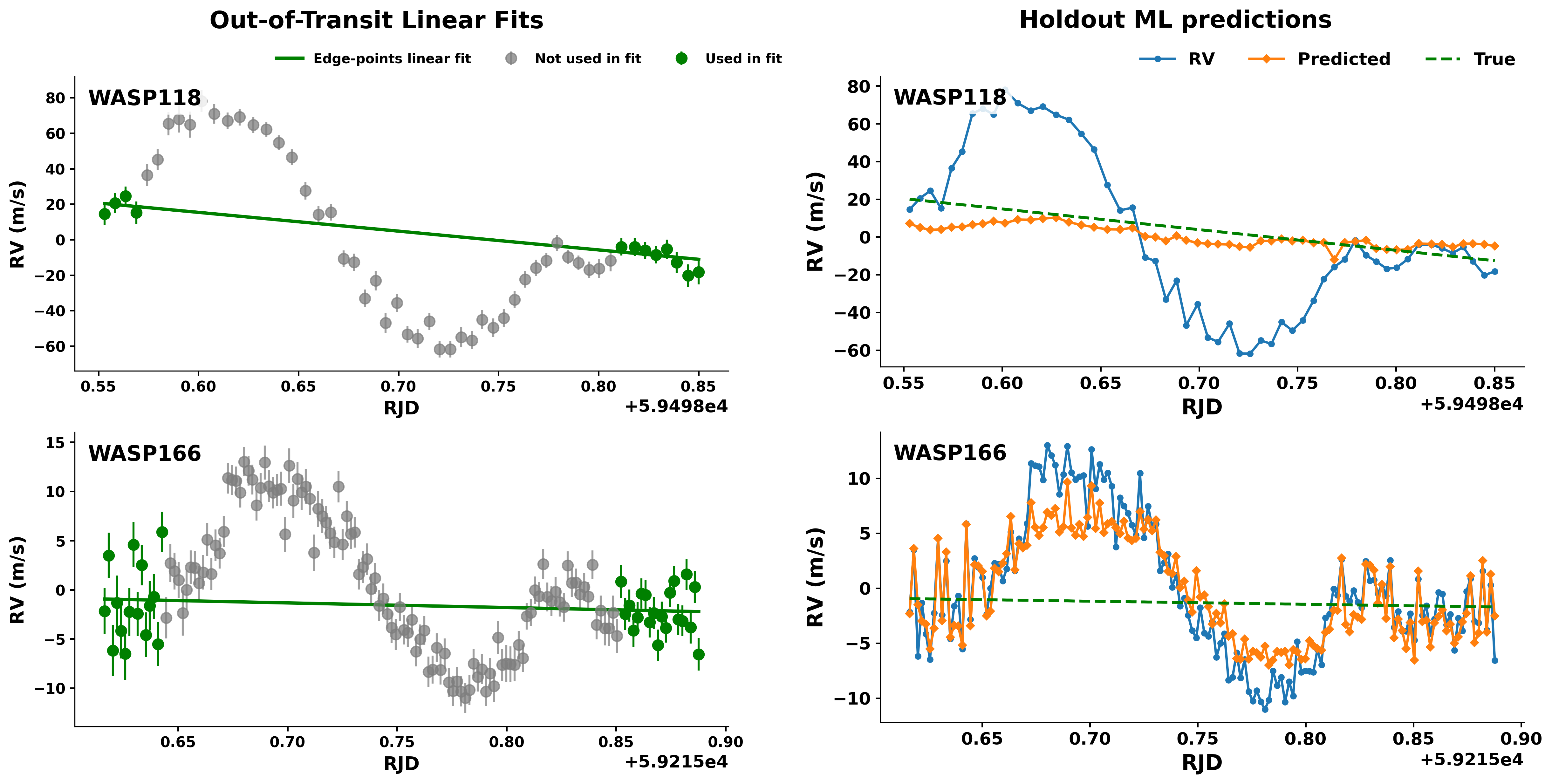}
\caption{
\textbf{Left:} RM time series for WASP-118 and WASP-166 after subtraction of the mean RV. Green points indicate the out-of-transit measurements used to estimate the reference RV trend (solid line), while grey points are affected by the RM anomaly and are excluded from the fit.
\textbf{Right:} Leave-one-star-out predictions for the same systems. The panels compare the observed RVs (blue), the ML-predicted reference RV trend (orange), and the reference RV trend derived from the out-of-transit fit (green).
}
\label{fig:rm_rv_fig}
\end{figure}

To define the regression target, we approximated the underlying RV evolution during each RM sequence by fitting a linear trend to the out-of-transit measurements obtained before ingress and after egress (Figure~\ref{fig:rm_rv_fig}, left panels). The fitted trend provides an estimate of the underlying RV evolution during the observing sequence in the absence of the RM anomaly. Ideally, this trend should be described by a Keplerian model; however, for the nearly circular systems considered here, the short transit durations relative to the orbital periods imply that the difference from a local linear approximation is much smaller than the RM signal. The mean-subtracted fitted trend was adopted as the reference target for the models.

\section{Machine-learning approach}

The ML models were trained to reconstruct the reference RVs. The input features included the mean-subtracted CCF-derived RVs together with the  line-profile diagnostics and activity indicators. Because different stars exhibit different RV amplitudes and activity levels, all features were normalised independently for each observing sequence. We explored mean-subtracted and fractional representations of the line diagnostics together with summary statistics of the activity indicators.

We evaluated Ridge regression, tree-based ensemble methods (Random Forests, Extremely Randomised Trees, XGBoost, LightGBM, and CatBoost), multilayer perceptrons, and voting and stacking ensembles. Hyperparameters were optimised using the Optuna framework.

Model performance was assessed using a leave-one-star-out strategy, where all observations of a given star were excluded from training and used exclusively for testing. The best overall performance, quantified by the smallest mean RMSE across the leave-one-star-out experiments, was obtained using a voting ensemble combining tree-based ensemble regressors. The selected feature set consisted of the mean-subtracted RVs, fractional variations of the line-profile diagnostics, and the RM sequence mean and standard deviation of the Ca activity indicator.

The observational uncertainties were not explicitly incorporated into the baseline training. We explored weighted training and bootstrap resampling based on the reported uncertainties, but the resulting improvements were marginal.

We also explored transfer learning using $\sim$100,000 synthetic solar spectra generated with SOAP simulations \citep[][]{Cristo-25}. The simulations included injected Doppler shifts together with spots and plages of different sizes and positions. After processing the simulated spectra through the same CCF pipeline as the observations, the models were pre-trained on the simulations and fine-tuned on the RM sample. No consistent improvement was obtained, likely because of the differences between the simulated activity signals and the observed RM data.

The predictive performance nevertheless varied significantly among the holdout stars (Figure~\ref{fig:rm_rv_fig}). Stars for which the RM-induced RV residuals (observed RV minus reference RV) correlated strongly with one or more line-profile diagnostics were generally well predicted by the ML models. In contrast, stars showing little or no correlation with the available diagnostics typically had weaker RM signals and larger relative uncertainties, resulting in poorer predictions. Moreover, Mahalanobis distances in feature space showed that several poorly predicted stars also lay outside the distribution of the training sample, requiring the models to extrapolate beyond the parameter space covered during training.

As an exploratory test, we applied the best-performing model to Sun-as-a-star observations (with HARPS-N) and to observations of Proxima Centauri (with ESPRESSO) retrieved from the DACE platform. For both targets, the Mahalanobis distances computed in the model feature space were substantially larger than those of the RM training sample, indicating that the predictions were being made far outside the domain covered by the training data. The reliability of the inferred corrections is therefore uncertain. Although the predicted RVs recovered periodicities consistent with known planetary signals, a simple generalized Lomb--Scargle analysis also revealed strong rotational modulation \citep[e.g.][]{Cameron-19, Mascareno-20}. We therefore find no compelling evidence that the current models can fully remove activity-induced RV signals in these systems.

\section{Conclusions}

In this proof-of-concept experiment, we demonstrated that part of the apparent RV variations arising from flux-induced line-profile distortions can be reconstructed using line-profile diagnostics derived from CCFs. Future improvements may come from larger and more diverse training samples, observations with smaller uncertainties, and applications restricted to stars with properties similar to those represented in the training set. Because the RM effect is fundamentally a flux-blocking phenomenon, it resembles the flux-imbalance component of spot-induced RV signals more closely than activity signals induced by brighter regions, such as faculae and plages, making spot-dominated stars more promising targets for future applications. The RM effect, however, does not reproduce RV variations caused by the magnetic suppression of convective blueshift.

\begin{acknowledgments}
This work was supported by  Funda\c{c}\~ao para a Ci\^encia e Tecnologia (FCT) through the research grant UID/04434/2025 (DOI 10.54499/UID/04434/2025).  V.A. acknowledges support from FCT CEEC program (reference 2023.06055.CEECIND/CP2839/CT0005, DOI: 10.54499/2023.06055.CEECIND/CP2839/CT0005).  N.C.S is co-funded by the European Union (ERC, FIERCE, 101052347).  K.A. acknowledges support from the Swiss National Science Foundation under the Postdoc Mobility grant P500PT\_230225.

\end{acknowledgments}



\bibliography{ref}{}
\bibliographystyle{aasjournalv7}



\end{document}